\documentclass[a4paper, 12pt]{article} 
\usepackage[right=1.1in,left=1.1in,top=1.1in,bottom=1.1in]{geometry}
\usepackage{graphicx} 
\usepackage{braket}
\usepackage{mathrsfs}
\usepackage[T1]{fontenc} 
\usepackage{amsmath}
\usepackage{pxfonts}
\usepackage[T1]{fontenc}
\usepackage{amssymb}
\usepackage{natbib}
\usepackage{epstopdf}
\usepackage{pifont}
\usepackage{setspace}
\usepackage{enumitem}
\usepackage{sectsty}
\usepackage{wrapfig} 
\sectionfont{\large}
\bibpunct{(}{)}{,}{a}{}{,}
\usepackage{tikz}   
\usepackage{tikz-3dplot} 

\newcommand{\qed}{\nobreak \ifvmode \relax \else
      \ifdim\lastskip<1.5em \hskip-\lastskip
      \hskip1.5em plus0em minus0.5em \fi \nobreak
      \vrule height0.75em width0.5em depth0.25em\fi}

\usepackage{bbm}
\usepackage{MnSymbol}
\usepackage[all]{xy}

\usepackage{xcolor,colortbl,array,amssymb}

\newcommand{\cmark}{\text{\ding{51}}}
\newcommand{\xmark}{\text{\ding{55}}}

\makeatletter

\title{\textbf{Realism about the Wave Function} } 

\author{Eddy Keming Chen\thanks{Department of Philosophy MC 0119,  University of California, San Diego, 9500 Gilman Dr, La Jolla, CA 92093-0119. Website: www.eddykemingchen.net. Email: eddykemingchen@ucsd.edu  }}

\date{Forthcoming in \emph{Philosophy Compass} \\ Penultimate version of June 12, 2019} 

\begin{document}
\bibliographystyle{plain}

\maketitle 



\begin{abstract}
A century after the discovery of quantum mechanics, the meaning of quantum mechanics still remains elusive. This is largely due  to the puzzling nature of the wave function, the central object in quantum mechanics.  If we are realists about quantum mechanics,  how should we understand the wave function? What does it represent? What is its physical meaning? Answering these questions would improve our understanding of what it means to be a realist about quantum mechanics.  In this survey article, I review and compare several realist interpretations of the wave function. They fall into three categories:  ontological interpretations,  nomological interpretations, and the \emph{sui generis} interpretation. For simplicity, I will focus on non-relativistic quantum mechanics. 

\end{abstract}

\hspace*{3,6mm}\textit{Keywords: quantum mechanics, wave function,  quantum state of the universe, scientific realism, measurement problem, configuration space realism, Hilbert space realism, multi-field, spacetime state realism, laws of nature}   

\begingroup
\singlespacing
\tableofcontents
\endgroup








\nocite{ney2013wave, SiderTOM, matarese2018challenge, allori2008common, allori2013primitive, albert1994quantum, maudlin2007completeness}

\section{Introduction}

Quantum mechanics is one of the most successful physical theories to  date. Not only has it been confirmed through a wide range of observations and experiments, but it also has led to technological advances of a breathtaking scale. From electronics and optics to computing, the applications of quantum mechanics are ubiquitous in our lives.

As much as it has given rise to technological innovations, the meaning of quantum mechanics remains elusive.  Many curious features of quantum mechanics, such as entanglement, non-locality, and randomness, are taken to be  \emph{prima facie} challenges for a clear understanding of quantum mechanics. These puzzles are related to the \emph{wave function}, the central object in quantum mechanics. Understanding the meaning of quantum mechanics seems to require a good understanding of the meaning of the wave function. 

What does the wave function represent?  That is the main concern of this survey article. The answer to that question is complicated by the fact that the wave function does not look like anything familiar. It is a function defined on a vastly high-dimensional space, with values in complex numbers, and unique only up to an ``overall phase.'' Nevertheless, we have devised many ways of using wave functions in making predictions and explaining phenomena. We use  wave functions to calculate the probabilities of  microscopic and macroscopic behaviors of physical systems. These led to the successful explanations of the double-slit experiment, the Stern-Gerlach experiment, and the stability of the hydrogen atom.  The wave function is indispensable for making these predictions. However, the predictions are probabilistic. (More on this later.) 

Roughly speaking,  there are three main views about the wave function:
\begin{description}
\item[Instrumentalism:] The wave function is merely an instrument for making empirically adequate predictions. 
\item[Epistemicism:] The wave function merely represents the observer's uncertainty of the physical situation.\footnote{The recently published theorem of \cite{pusey2012reality} shows that a certain class of epistemic interpretations of the wave function are incompatible with the empirical facts.} 
\item[Realism:] The wave function represents something objective and mind-independent. 
\end{description} 
In this article, I focus on the realist interpretations of the wave function. They seem to be the most interesting and promising ways of understanding quantum mechanics. 

Let me make four remarks.  First, the meaning of the wave function is related to solutions to the quantum measurement problem. Hence, we will start in \S 2 with an introduction to that problem, along with some mathematical preliminaries. Second, I  left the definition of \emph{realism} open-ended. That is because we will consider proposals for specific versions of realism about the wave function. The proposals are grouped into three  categories:  ontological interpretations (\S 3),  nomological interpretations (\S 4), and the \emph{sui generis} interpretation (\S 5). Third, because of the prevalence of quantum entanglement, ``the wave function'' should be understood to refer to the wave function of the universe, or the universal wave function. The wave functions of the subsystems are thought to be derivative of the universal one. Fourth, for simplicity, I will focus on non-relativistic versions of quantum mechanics.\footnote{For complications that arise in the relativistic theories, see \cite{myrvold2015wavefunction}. } 

The issues taken up here are continuous with the general question about how to interpret physical theories. They offer concrete case studies for scientific realism, and they might be useful for  philosophers of science, metaphysicians, and anyone with an interest in understanding quantum mechanics. 

\section{Background } 

In this section, we will review some basic facts about the wave function and its connection to  probabilistic predictions. We will then consider the quantum measurement problem and three realist theories that solve it. The upshot is that the wave function occupies a central place in their descriptions of physical reality. 

\subsection{The Wave Function}

It will be useful to have a brief review of classical mechanics. To describe a classical mechanical system of $N$ particles, we can specify the (three-dimensional) position $\boldsymbol{q_i}$ and  the (three-dimensional) momentum $\boldsymbol{p_i}$ of each particle in physical space (represented by $\mathbb{R}^3$). We can represent the classical state of an $N$-particle system in terms of $6N$ numbers, $3N$ for positions and $3N$ for momenta. The classical state can also be represented as a point in an abstract state space called the \emph{phase space} $\mathbb{R}^{6N}$.  Once we specify the forces (or interactions) among the particles,  they evolve deterministically, by the Hamiltonian equations of motion: 

\begin{equation}\label{HE}
\frac{\partial \boldsymbol{q_i}}{\partial t} = \frac{\partial H}{\partial \boldsymbol{p_i}} \text{  ,  } \frac{\partial \boldsymbol{p_i}}{\partial t} = - \frac{\partial H}{\partial \boldsymbol{q_i}},
\end{equation}
where $H$ stands for the Hamiltonian function on the  phase space, which is a short-hand notation that encodes classical interactions such as Newtonian gravitational potential and Coulomb electric potential. The Hamiltonian equations are differential equations, and the changes in the particles are obtained from taking suitable derivatives of $H$.  In this sense, $H$ is the generator of motion. For every point in  phase space, $H$ generates a curve starting from that point. In other words, for every initial condition of the $N$ particle system, $H$ determines the future trajectories of the particles. 

Now let us introduce the quantum mechanical way of describing a system of $N$ ``particles.''\footnote{In some ways of thinking about quantum mechanics, particles are not fundamental. Hence the quotation marks. } Instead of describing it in terms of the positions and the momenta of $N$ particles, we use a wave function for the system. The wave function represents the \emph{quantum state} of the system. In the position representation, the wave function, denoted by $\psi(q)$, is a particular kind of function from  configuration space $\mathbb{R}^{3N}$ to complex numbers $\mathbb{C}$. Let us elaborate on this definition: 

\begin{itemize}
\item Domain: the domain of the wave function $\psi$ is $\mathbb{R}^{3N}$, or $N$ copies of physical space $\mathbb{R}^{3}$. $N$ is the total number of particles in the system. When $N$ is large,  $\mathbb{R}^{3N}$ is vastly high-dimensional. Each point in $\mathbb{R}^{3N}$ is an $N$-tuple $(\boldsymbol{q_1}, ..., \boldsymbol{q_N})$. Each  $\boldsymbol{q_i}$ corresponds to particle $i$'s position in physical space $\mathbb{R}^{3}$. Hence, the $N$-tuple lists the positions of $N$ particles. We use a point in $\mathbb{R}^{3N}$ to represent a particular configuration (arrangement) of $N$ particles in $\mathbb{R}^{3}$. Hence, $\mathbb{R}^{3N}$ is called the configuration space.\footnote{This is the ordered configuration space, in which a permutation of the particle labels creates a different configuration. If the particles are indistinguishable, then it is more natural to use the \emph{unordered configuration space}, $^N\mathbb{R}^{3}$. This has implications for the nature of the wave function. See \cite{ChenOurFund} and the references therein.}   The wave function $\psi(\boldsymbol{q_1}, ..., \boldsymbol{q_N})$ is a function whose domain is the configuration space, which is vastly high-dimensional when the system has many particles. 


\item Range: the range of the wave function $\psi$, in the simplest case, is the field of complex numbers $\mathbb{C}$. A complex number has the form $a+bi$, where $i=\sqrt{-1}$; in polar form, it is $Re^{i\theta}$, where $R$ is the amplitude and $\theta$ is the phase.\footnote{If we include spinorial degrees of freedom, the range is $\mathbb{C}^k$. We set spins aside in this paper.}

\item Restrictions: the wave function is a particular kind of function from $\mathbb{R}^{3N}$ to $\mathbb{C}$. It has to be a ``nice''  function that we can take certain operations of integration and differentiation.\footnote{It has to be ``square-integrable.'' That is, if we take the square of the amplitude of the wave function value at every point, and integrate over the entire configuration space, we will get a finite value. This is to ensure that we can normalize the squared value of the wave function to $1$ so that it has meaningful connections to probabilities. To ensure that we can take suitable derivatives on the wave function, we often also require the wave functions to be sufficiently \emph{smooth}.} 

\item Abstract state space: each wave function describes a quantum state of the system. The space of all possible quantum states is called the \emph{state space} of quantum mechanics. The state space will include all possible wave functions for the system, that is, all the ``nice'' functions from configuration space $\mathbb{R}^{3N}$ to complex numbers $\mathbb{C}$.  Hilbert space is the abstract mathematical space that we use to describe such a state space.  Hilbert space can be a high-dimensional vector space, in which each wave function is represented as a vector. 
\end{itemize}

In classical mechanics, the state of a system is represented by the positions and the momenta of  $N$ particles (a point in phase space) that change deterministically according to (\ref{HE}).  If the wave function represents the quantum state of a system at a time, how does it change over time? It obeys another differential equation called the \emph{Schr\"odinger equation}: 
\begin{equation}\label{SE}
 i\hbar \frac{\partial \psi}{\partial t} = \hat{H} \psi,
\end{equation}
where $i$ is the complex number $\sqrt{-1}$, $\hbar$ is the Planck constant divided by $2\pi$, and $\hat{H}$ is the Hamiltonian operator that encodes the energy and fundamental interactions in nature. It is also deterministic: given any vector in  Hilbert space, the Schr\"odinger equation (\ref{SE}) produces a determinate curve in  Hilbert space.  Another  feature of  (\ref{SE}) is that it is linear: if $\psi_1$ and $\psi_2$ are solutions to the equation, then their linear combinations are also solutions to the equation. A surprising consequence of linearity is that, in the  Schr\"odinger's cat thought experiment, the system can be in a ``superposition'' of two macroscopically distinct states: 
\begin{equation}\label{cat}
\Psi = \frac{1}{\sqrt{2}}\psi_{alive} + \frac{1}{\sqrt{2}} \psi_{dead}
\end{equation}
A cat  in this quantum state is not  alive, and it is not  dead. The linear Schr\"odinger equation (\ref{SE}) ensures that the wave function of the system will not change into $\psi_{alive}$ (the cat is  alive and the atom has not decayed) or $\psi_{dead}$ (the cat is dead and the atom has decayed). Thus, the Schr\"odinger equation does not determine a unique  experimental outcome. To resolve this,  textbook quantum mechanics supplements the Schr\"odinger equation with additional collapse postulates. Whenever we open the box and ``observe'' the cat, the system will suddenly change (collapse) into one of the two states: $\psi_{alive}$  or $\psi_{dead}$. An important role of the wave function is determining the probabilities of experimental outcomes, which are taken to be the results of wave function collapses. For example, the probability of finding the system in any set of configurations is given by the Born rule:
\begin{equation}\label{Born}
P( q \in A) = \int_{A} |\psi(q)|^2 dq,
\end{equation}
where $A$ is a set of points in  configuration space, $|\psi(q)|^2$ is the  squared  amplitude of the wave function, and  $dq$ is the Lebesgue measure on $\mathbb{R}^{3N}$.  In the cat example,  the probability of finding the cat to be alive is equal to $\frac{1}{2}$, since $ \int |\frac{1}{\sqrt{2}}\psi_{alive} |^2 + 0 = \frac{1}{2}$. The Born rule has the consequence that wave functions that differ only by an \emph{overall phase} (multiplied by a complex number $e^{i\theta}$, where $\theta \in [0,2\pi]$) will give rise to the same observable phenomena ($|\psi|^2 = |e^{i\theta}\psi|^2$). That is called the \emph{overall phase symmetry}, which motivates the common view that two wave functions that differ by an overall phase  represent the same quantum state. 

\subsection{The Quantum Measurement Problem}

Notwithstanding the empirical success of quantum mechanics,  the collapse postulates introduce a host of difficulties.  If the wave function (of the system and the measurement device) obeys the Schr\"odinger equation, how can it also obey the collapse postulates that contradict the linearity of the Schr\"odinger equation? But if the wave function does not collapse, how can we obtain unique experimental outcomes? In short, we have the \emph{quantum measurement problem}: 

\begin{itemize}
\item[(P1)] The wave function is the complete description of the physical system.
\item[(P2)] The wave function always obeys the Schr\"odinger equation.
\item[(P3)] Every experiment has a unique outcome. 
\end{itemize}
Each of these three propositions is, on its own, plausible. However, together they lead to a contradiction. To see the contradiction,  let us apply them to Schr\"odinger's cat thought experiment. If P1 is true,  the system is completely described by (\ref{cat}). If P2 is true,  the wave function never collapses into one of the definite states. If P3 is true, the cat is nonetheless in one of the definite states---either alive or dead.\footnote{For a more thorough discussion about the quantum measurement problem, see \cite{sep-qt-issues}, \cite{albert1994quantum}, and \cite{bell1990against}. } 

Since P1---P3 are inconsistent,  at least one of them is false. Rejecting P1 or P2 would require us to develop alternative theories of quantum mechanics, since we would need to find additional variables omitted by the wave function, or we would need to modify the Schr\"odinger equation. Rejecting P3 would lead to major revisions of our assumptions about the world. There are three main  ``interpretations'' of quantum mechanics that carry out such strategies. They all contain significant revisions of quantum mechanics, so we should call them realist theories of quantum mechanics instead of interpretations. 

First, the de Broglie-Bohm theory, or Bohmian mechanics (BM), rejects P1. According to BM, the wave function is not the complete description of the physical system. There are actual particles with precise positions in physical space. The wave function still obeys the Schr\"odinger equation. But the wave function also determines the velocity of the particles according to the \emph{guidance equation}.\footnote{ The particles move according to the guidance equation:
\begin{equation}\label{GE}
 \frac{dQ_i}{dt} = \frac{\hbar}{m_i} \text{Im} \frac{ \nabla_i \psi }{  \psi}, 
\end{equation}
where $Q_i$ and $m_i$ are the position and mass of particle $i$, Im means taking the imaginary part, and $\nabla_i $ means taking the gradient with respect to the $i$-th particle. The particles are initially distributed according to the Born rule, and their distribution will always agree with the Born rule because of the mathematical properties of the Schr\"odinger equation and the guidance equation.}  In the cat example, the cat is made out of particles in physical space. There is always a determinate configuration of particles, so the cat is either alive or dead. The probabilities of quantum mechanics become (more or less) epistemic uncertainties over initial particle configurations.\footnote{ For a survey of BM, see \cite{sep-qm-bohm}; for the original paper, see \cite{bohm1952suggested}; for a modern version, see \cite{durr1992quantum}.}

Second, the Ghirardi-Rimini-Weber theories of spontaneous collapse (GRW) reject P2. According to GRW theories, the wave function $\psi_{t}$ does not always obey the Schr\"odinger equation. It undergoes spontaneous collapses with a fixed rate per particle per unit time. In the cat experiment, given the vast number of particles in the system, it will quickly collapse into a determinate state in which the cat is either alive or dead.  Collapses are represented by Gaussian functions with a fixed width in physical space. Due to entanglement, collapses on a single particle has the effect that the universal wave function will collapse into a definite state.\footnote{Representing collapses by Gaussian functions gives rise to the ``tails problem.'' See \cite{sep-qm-collapse} \S12 and the references therein. } On the macroscopic scale, the collapse will give rise to (approximately) Born rule probabilities. Each version of GRW postulates specific values for the collapse rate and the Gaussian width. Moreover, there can be additional variables representing ontology in physical space. GRWm adds a mass-density ontology that specifies the amount of mass in physical space by a real-valued function $m(x,t)$, where $(x,t)$ is a space-time point.\footnote{The mass-density function is defined from the wave function: 
\begin{equation}\label{mxt}
m(x,t) = \sum_{i=1}^{N} m_{i} \int_{\mathbb{R}^{3N}} d^{3} x_{1} ... d^{3} x_{N} \delta^{3} (x_{i} - x) |\Psi_{t}(x_{1},..., x_{N})|^2 
\end{equation}
}  
In contrast, GRWf adds a flash ontology that postulates the existence of space-time events at the center of the Gaussian function. It can be represented as a function $F(x,t)$ with $x\in  \mathbb{R}^{3}$ and $F(x,t) = 1$ if $(x,t)$ is the center of some GRW collapse and $0$ otherwise.\footnote{For a survey of GRW, see \cite{sep-qm-collapse}; for the original paper, see \cite{ghirardi1986unified};  \cite{bell2004speakable}, Ch 22, contains a clear presentation of the theory. } 

Third, many-worlds interpretations of Everettian quantum mechanics (EQM) reject P3.\footnote{The many-worlds interpretations are popular among Everettians. However, \cite{conroy2012relative} has provided textual evidence that it is possible to understand Everett as endorsing a single-world interpretation of quantum mechanics.  } According to these interpretations, there is no need to ensure that there is a unique outcome in the cat experiment. There simply are two branches of the wave function, one in which the cat is alive and the atom has not decayed and the other in which the cat is dead and the atom has decayed. Both branches co-exist. Because of a property called \emph{decoherence}, the branches do not interfere much with each other. The branches of the wave function are emergent worlds, the wave function is the complete description of the ``multi-verse,'' and it always obeys the Schr\"odinger equation. Similarly to GRWm, we can devise a version of EQM with a mass-density ontology. This is called Sm and was first proposed by \cite{allori2010many}.  A challenge for any version of EQM is how to make sense of probability in a world in which every possible outcome of every quantum experiment  happens with certainty.\footnote{For a survey of EQM, see \cite{sep-qm-manyworlds}; for the original paper, see \cite{everett1957relative}; for an updated book-length development of the theory, see  \cite{wallace2012emergent}. There has been significant progress in addressing the probability challenge with the tools of typicality, decision theory, and self-locating probabilities. For some recent examples, see \cite{barrett2017typical},  \cite{wallace2012emergent},  \cite{sebens2016self}, and the references therein. } 
 
The upshot  is that the wave function figures prominently in all three realist quantum theories. In BM, although the wave function is not the complete description of the system, it is still part of the description. Moreover, the wave function guides particle motion. In GRW,  the wave function collapses and gives rise to unique outcomes of experiments. In (many-worlds interpretations of) EQM, the wave function never collapses but  gives rise to emergent parallel worlds. For quantum theories with additional ontology, such as BM, GRWm, GRWf, and Sm, the wave function is also tied to the dynamics of the additional ontology. But their relationship is different in these theories.  Bohmian particles have independent dynamics: even if the wave function were not to change,  Bohmian particles would still move in a non-trivial fashion. That is not the case in GRWm, GRWf, and Sm. Had there been no change to the wave function, the additional ontology would not change either. It is in this sense that the dynamics of mass-densities and flashes are not independent of the dynamics of the wave function. 

\section{Ontological Interpretations}

In this section, I review four \emph{ontological} interpretations of the wave function. However, the label ``ontological'' could be misleading. These four interpretations share the feature that the wave function is interpreted as part of the fundamental \emph{material ontology}, on a par with particles, fields, space-time events or properties, which are the kind of microscopic things that make up macroscopic objects such as tables and chairs. In \S 4 and \S 5, we will review  nomological interpretations and the \emph{sui generis} interpretation of the wave function,  which  are compatible with the position that the wave function is part of the ontology but just not in the same ontological category as particles or fields.

\subsection{A Field on a High-Dimensional Space}

According to the first ontological interpretation, the fundamental space is a high-dimensional space, and the wave function is a field in that space. This was introduced by \cite{AlbertEQM}. Albert calls this view \emph{wave function realism}.\footnote{It has been developed and defended by  \cite{LoewerHS}, \cite{ney2012status, ney2013ontological}, and \cite{NorthSQW}, although  North is primarily concerned with the first part of the thesis, i.e. the fundamental space is a high-dimensional space.} However, as we shall see in the later sections, that label is no longer appropriate given the abundance of other approaches that are also realist about the wave function. 

It is counterintuitive how the fundamental space can be high-dimensional. It might help to compare this idea with something familiar---classical physics.  In classical field theories such as Maxwellian electrodynamics, electromagnetic fields are fields on the four-dimensional physical space-time.  A field on physical space-time can be interpreted as an assignment of monadic properties (field strength and direction) to each point in space-time. Such an assignment is contrained by  Maxwell's equations and certain boundary conditions.

In a similar way, the wave function can be interpreted as a physical field. However, the wave function cannot be interpreted as a field on physical space, as its domain is the high-dimensional configuration space, represented by $\mathbb{R}^{3N}$. If we take   configuration space to be the fundamental space, then the wave function can be interpreted as a field that assigns properties to each point in  configuration space. The properties assigned by the field, represented as complex numbers, change according to the  Schr\"odinger equation. On this view, the high-dimensional configuration space is ontologically prior to  physical space(time), and the latter somehow comes out of the fundamental structure. This is despite the fact that we call the high-dimensional space ``configuration space,'' which seems to imply the reverse order of ontological dependence.\footnote{There are three versions of this view:\begin{itemize}\item Bohmian version: the fundamental space is represented by $\mathbb{R}^{3N}$. The fundamental ontology consists in a point particle located in that space and a field that assigns properties to points of that space. The field always evolves by the Schr\"odinger equation. The point particle moves along in the field according to the guidance equation, much like corks move along in a  river. Here we see a dis-analogy with the classical field. In classical physics, the field and the particles satisfy the action-reaction principle; the fields and the particles can influence each other. In Bohmian mechanics, the wave function interpreted as a field can influence the particles but not the other way around. 
\item GRW version: the fundamental space is represented by $\mathbb{R}^{3N}$. The fundamental ontology consists in a field that assigns properties to points of that space. The field evolves by the Schr\"odinger equation most of the time but sometimes collapses by the GRW collapse mechanism. 
\item Everettian version: the fundamental space is represented by $\mathbb{R}^{3N}$. The fundamental ontology consists in a field that assigns properties to points of that space. The field always evolves by the Schr\"odinger equation. 
\end{itemize}
The high-dimensional field interpretation of the wave function is incompatible with GRWm, GRWf, or Sm. 
} 

The high-dimensional field interpretation prioritizes the structure of the wave function and its dynamics. The fundamental physical events are those that happen on the high-dimensional space. A key challenge to this view is  to explain our apparent experiences in a three-dimensional space. That is not just a question about recovering the manifest image, but it is also about whether such an interpretation of quantum mechanics can be ``empirically coherent,'' in the sense that if our evidence for quantum mechanics comes from instrument readings in the three-dimensional space, the theory should not undermine such evidence. It should explain how the appearances of three-dimensional objects come out of the high-dimensional fundamental space.\footnote{See \cite{barrett1999quantum} and \cite{barrett1996empirical}.} 

Albert (1996) suggests that the explanation lies in the dynamics---in the structure of the  Hamiltonian operator. Although all the $3N$ dimensions are metaphysically on a par:
\begin{equation}
\{q_1, q_2, q_3,  q_4, q_5, q_6, ... , q_{3N-2}, q_{3N-1}, q_{3N} \}
\end{equation}
 the Hamiltonian operator has a term that encodes fundamental interactions and it takes on a particular form: 
\begin{equation}
\mathop{\sum\sum}_{1 \leq i<j \leq N} V_{ij} [ (q_{3i-2} -  q_{3j-2})^2 + (q_{3i-1} - q_{3j-1})^2 +(q_{3i} - q_{3j})^2 ] 
\end{equation}
The Hamiltonian operator groups the coordinates in the $3N$-dimensional configuration space into triplets, such that there might be emergent objects that have the same functional profile as what we take to be ordinary objects in the 3-dimensional space. This provides reasons to believe that there might be an emergent $3$-dimensional physical space. For a discussion of Albert's proposed explanation of emergence, see  \cite{monton2002wave}, \cite{lewis2004life},  \cite{ney2012status}, \cite{ChenOurFund, emery2017against}, and \cite{ney2017finding}. 

\cite{maudlin2013nature} criticizes the high-dimensional field interpretation on the ground that it reifies redundant structure. The common view (\S 2.1) holds that two wave functions that differ only by a  multiplicative constant represent the same physical state.  But if we interpret the wave function as a field that assigns numbers (that represent physical magnitudes) to points in  configuration space, then we would distinguish two wave functions related by a constant, for the numbers assigned to the points are different.\footnote{This problem can be avoided if we adopt  an \emph{intrinsic} (or gauge-free) characterization of the wave function, in terms of comparative relations that are invariant under any change by a multiplicative constant. On the comparativist view, the fundamental structure consists in not monadic properties but comparative relations holding between points of the high-dimensional space. See \cite{chen2017intrinsic} for a proposal. }

\subsection{A Multi-field on Physical Space}

The high-dimensional field interpretation of the wave function faces difficulties, primarily because it privileges  configuration space over physical space. There are many good reasons to take physical space to be ontologically more basic. First, it underlies many important symmetries in physics. Second, it is much easier for a theory to be empirically coherent if it does not undermine the relative fundamentality of physical space(time). 

These difficulties are avoided in the second ontological interpretation, according to which the fundamental space is the ordinary physical space(time). On this view, the wave function is not a field in the traditional sense, but a \emph{multi-field} on physical space. (See \cite{forrest1988quantum}, \cite{belot2012quantum}, \cite{chen2017intrinsic, ChenOurFund}, \cite{Hubert2018}.) A multi-field is similar to a field. However, unlike fields, multi-fields assign properties not to individual points but to \emph{regions} of points in space.  Such regions can be connected or disconnected. Mathematically, the wave function is a function from $N$ copies of $\mathbb{R}^{3}$ to complex numbers. Instead of thinking of it as a field that assigns properties to every point in $\mathbb{R}^{3N}$, we can think of it as a ``multi-field'' that assigns properties to every region of $\mathbb{R}^{3}$ that is composed of $N$ points. The multi-field interpretation is a more faithful representation  for ``indistinguishable particles,'' for which particle labels do not matter. This is because spatial regions understood as $N$-element subsets of  $\mathbb{R}^{3}$ (or mereological fusions of $N$ points in $\mathbb{R}^{3}$) are \emph{unordered}. Thus, the multi-field interpretation has  the additional advantage of automatically enforcing what is called ``permutation invariance'': mere permutations of a configuration of $N$ particles do not change the physical state.

\subsection{Properties of Physical Systems}

The third ontological interpretation, proposed by \cite{wallace2010quantum}, affirms the (relative) fundamentality of the physical space(time). On this view, we assume that the universe has a compositional structure, i.e. it is divided into subsystems that occupy some spatial-temporal regions.\footnote{This assumption is important and can turn out to fail in some theories. For the Everettian interpretation according to which the quantum state or the wave function represents everything there is, the compositional structure may need to be emergent and is certainly not fundamental. See \S3.4 for a discussion of that possibility.} Larger systems can be made out of unions of smaller systems. And the universe is the union of all systems.  Although not every system has a wave function (because of entanglement), we can still associate to each system a determinate property represented by what is called a \emph{density matrix}. A density matrix encodes much dynamical information of the system.\footnote{For example, if the universe consists in two systems $A$ and $B$, and if their joint quantum state is  this:
\begin{equation}
\ket{\Psi_{AB}} = \frac{1}{\sqrt{2}} (\ket{\alpha}_A \ket{\beta}_B - \ket{\beta}_A \ket{\alpha}_B)
\end{equation}
then the density matrix associated with system $A$ (called the \emph{reduced density matrix}) will be: 
\begin{equation}
\rho_{A} = \frac{1}{2} (\ket{\alpha}_A\bra{\alpha}_A  +  \ket{\beta}_A \bra{\beta}_A)
\end{equation}
On this view,  $\ket{\Psi_{AB}}$ and $\rho_{A} $ are understood  as representing intrinsic properties of the systems or the space-time locations of the systems.
} 
This view was introduced as an alternative to the high-dimensional field interpretation. It is also an alternative to the low-dimensional multi-field interpretation. However, it is still a version of realism about the wave function, since the universal wave function (or the universal density matrix) is to be found in the ontology---the property of the entire universe.

Wallace and Timpson call this approach \emph{spacetime state realism}. Compared to the field interpretations in \S3.1 and \S3.2,  spacetime state realism seems more minimal, in the sense that it is neutral with respect to the specific structures of quantum theories. As long as the quantum theory comes with a compositional structure (a decomposition of larger systems into smaller systems), we have a well-defined procedure of ascribing properties to physical systems.  They argue that  it also has significant advantages in handling  relativistic invariance. See  \cite{Swanson2018f} for some discussions about the relativistic extensions. 

A question about spacetime state realism is whether the fundamental ontology contains redundant structure. If we help ourselves to a decomposition of the universe into subsystems, and if we have the quantum state of the universe, then we can obtain density matrices of the subsystems by a purely mathematical procedure of tracing out the environmental degrees of freedom. Since they can be derived from the quantum state of the universe, the properties of the subsystem need not be placed in the fundamental ontology. Hence, spacetime state realism is in tension with the principle of parsimony that we should minimize postulating more fundamental structure than we need.  If we get rid of the subsystem properties and only keep the universal property (the universal density matrix), then this approach would be in the same spirit as the low-dimensional multi-field approach in \S3.2. However, if one is interested in finding the most \emph{perspicuous} interpretation (as suggested by Wallace and Timpson), and not the adequate ontology that is the most parsimonious and simple,  perhaps space-time state realism is the right choice. On the other hand, one might object that an ontology of properties represented by density matrices, which is not metaphysically impossible, may be much less perspicuous than an ontology of properties represented by numbers and vectors.\footnote{See \cite{monton2006quantum} and \cite{monton2013against} for another view that interprets the wave function as properties of physical systems.}

\subsection{A Vector in  Hilbert Space}

The final ontological interpretation of the wave function takes the abstract Hilbert space  more seriously.  Recall that the wave function is represented as a vector in  Hilbert space, and  the  Schr\"odinger equation can be represented as an equation for vector rotation in that space.  \cite{carroll2018mad} suggest that  Hilbert space is the fundamental space, and the wave function is just a vector in that space. Physical goings-on in the world will correspond to some particular directions the vector is pointing at. 

Since the Everettian interpretation of QM is the most natural place for this view, \cite{carroll2018mad} call this approach \emph{Mad-Dog Everettianism}. In their words, the label is ``to emphasize that it is as far as we can imagine taking the program
of stripping down quantum mechanics to its most pure, minimal elements.''

It is already difficult to recover ordinary objects from  configuration space. It is even more difficult to recover them from  Hilbert space. For one thing, there is no space-time structure in  Hilbert space. The state of the world corresponds to a vector, which is just like every other vector. How can anything familiar, such as space, time, and ordinary objects, come out of a vector in a high-dimensional Hilbert space? Like Albert (1996), Carroll and Singh propose  that the answer lies in the structure of the Hamiltonian operator. The Hamiltonian provides a privileged way to decompose the total Hilbert space into smaller spaces, which may explain the emergent structure.\footnote{Their analysis is restricted to \emph{locally finite-dimensional Hilbert spaces}. See \cite{cotler2017locality} and \cite{bao2017hilbert} for more details.} Carroll and Singh's proposal is more speculative than the high-dimensional field interpretation, since  Hilbert space is more abstract than  configuration space. However, it is in part motivated by the non-fundamentality of space-time in several theories of quantum gravity. As such, it could be a  fruitful project to explore.

\section{Nomological Interpretations}

According to the previous ontological interpretations, the wave function is part of the fundamental material ontology of the world. The wave function in quantum mechanics is given the same status as particles and fields in classical mechanics. In contrast,  nomological interpretations hold that the wave function is nomological, i.e. on a par with laws of nature.  In this section, I survey two kinds of nomological interpretations of the wave function: the strong nomological interpretations and the weak nomological interpretations. 

These interpretations are most compelling from a Bohmian point of view. However, they might be adaptable for some versions of GRW theories and Everettian theories with additional ontologies. 

\subsection{Strong Nomological Interpretations}

The guiding idea of nomological interpretations is that the wave function is on a par with laws of nature (\cite{goldstein1996bohmian, goldstein2001quantum, goldstein2013reality}). To appreciate the strong nomological interpretations, it would be helpful to review the status of the Hamiltonian function in classical mechanics. As mentioned in \S 2.1, the Hamiltonian equations  (\ref{HE}) govern the motion of classical particles in physical space, represented by a curve in  phase space. The Hamiltonian function is the generator of such motion. It encodes the total energy of the system. In one of the more familiar cases, it is a short hand for the kinetic energy term and the pair-wise interactions of the particles (including gravitational, electric, and magnetic interactions).  We can write out $H$ explicitly as a function (of position and momentum) on the right hand sides of the equations. For the Hamiltonian equations to be simple \emph{laws} of nature, $H$ has to be a simple function. In this sense, we give $H$ a nomological interpretation. Although it is a function on phase space, we do not treat it as part of the material ontology.

Let us now consider Bohmian mechanics. In this theory, the guidance equation governs the motion of   particles in physical space, represented by a curve in  configuration space. The wave function is the generator of such motion. Moreover, there is no back reaction from the particles to the wave function. If the wave function turns out to be a simple function that is somehow fixed by the theory, then we can write out $\Psi$ explicitly as a function (of configuration variables) on the right hand side of the equation. In that case, we can give it an analogous nomological interpretation. Although it is a function on configuration space, we can treat it as not part of the ontology but only part of the law system. I call this the \emph{strong nomological interpretation}, for it affords the same status to the wave function as it does to the classical Hamiltonian function.  Just as both Humeans and non-Humeans about laws of nature can embrace the nomological interpretation of the Hamiltonian function in classical mechanics, both Humeans and non-Humeans can embrace the strong nomological interpretation of the wave function. 

However, the analogy between the Hamiltonian function in classical mechanics and the wave function in Bohmian mechanics is not perfect. While the Hamiltonian function is time-independent (it does not change over time), the wave function is typically time-dependent (it changes over time). For something to be nomological or to be like a fundamental law, one might expect it to be time-independent. However, there are  reasons to be optimistic. \cite{goldstein1996bohmian, goldstein2001quantum}, and \cite{goldstein2013reality} have offered one. When we eventually extend quantum mechanics to quantum gravity,  the wave function of the universe may turn out to be time-independent. This is seen in the Wheeler-DeWitt equation of canonical quantum gravity: 
\begin{equation}\label{WD}
 \hat{H} \Psi = 0
\end{equation}
If we understand (\ref{WD}) as telling us about the time evolution of the wave function, then it tells us that the wave function does not change over time, i.e. it is time-independent.\footnote{ Since the Schr\"odinger equation governs how the wave function changes over time, it is to be treated not as a fundamental equation but only as an effective equation---describing the behavior of subsystems.} It is worth emphasizing that a time-independent universal wave function does not entail that there is no change in the universe (cf: the problem of time in quantum gravity). In the Bohmian theory, given a time-independent  wave function, the particles can still move in a non-trivial way, since  a time-independent wave function can  generate a non-trivial velocity field in configuration space. 

The strong nomological interpretation faces some challenges. First, it is an open question whether the Wheeler-DeWitt equation governs the universal wave function. For example, there are research programs in quantum gravity that do not presuppose it. Second, since the wave function does not change over time, it requires some revisions about how we think about the arrow of time.\footnote{The problem is that in standard Boltzmannian quantum statistical mechanics, the arrow of time is associated with the increase of entropy of the quantum system, which is a property of the wave function. If the universal wave function is time-independent, then there is no increase of Boltzmann entropy. Perhaps the Bohmian approach can help by providing an alternative definition of entropy (or ``effective entropy'') in terms of particle configurations. } 

Why should fundamental laws be time-independent? One might appeal to one's intuition that laws are eternal and unchanging. But I think a better reason is that time-independent laws are more likely to be simple, and simplicity is the more fundamental consideration.\footnote{Here I am supposing that the demand for simplicity of the fundamental laws is neutral between Humeanism and non-Humeanism about laws of nature. For Humeans, theoretical virtues such as simplicity and informativeness (and the balance of the two) are constitutive of some facts being Humean laws of nature (that are the best summaries of the Humean mosaic). For non-Humeans, simplicity and informativeness are our epistemic guides to discover which facts are non-Humean laws. Both parties, it seems to me, should respect the pre-theoretical intuitions that fundamental laws should be as simple as possible. And the complexity of some principle is a serious strike against it being a fundamental law. } A time-dependent function may be more complicated than a time-independent one. And if a wave function $\Psi$ is time-independent, as suggested in (\ref{WD}),  then  $\Psi$ presumably has many symmetries, so that the different contributions under time-evolution will cancel out. Such symmetries \emph{may} ensure that the wave function is simple. For example, a translationally invariant function on physical space can only be a constant function, which is relatively simple.\footnote{Relatedly, \cite{allori2017new} has proposed a new argument for  nomological interpretations based on symmetry principles.} However, if simplicity is the more fundamental desideratum, we may consider whether time-independence is only a defeasible guide and whether there may be simple yet time-dependent laws. 

Can we find a simple (but maybe time-dependent) quantum state that plays the same role as the classical Hamiltonian? 
An example can be found in \cite{chen2018IPH}. Chen proposes a  choice of the initial quantum state that is as simple as the choice of the low-entropy macrostate specified in the Past Hypothesis (Albert 2000). The Past Hypothesis has been taken to be a candidate fundamental law of nature.\footnote{See discussions in \cite{ feynman2017character, albert2000time, callender2004measures, LoewerCatSLaw, wallace2011logic, wallace2012emergent} and \cite{loewer2016mentaculus}.}  If the Past Hypothesis is simple enough to be a law, then Chen's quantum state is simple enough to be nomological. However, in that approach, the fundamental quantum state is a mixed state rather than a pure state  (it has to be represented by a density matrix rather than a wave function). Moreover, the generator of motion becomes time-dependent. 


The strong nomological interpretation is most compelling in the Bohmian framework. However, In Everettian and GRW theories with additional ontologies, it may also be possible to make a case for the strong nomological interpretation.

\subsection{Weak Nomological Interpretations}

The literature on the nomological interpretation of the wave function is expanding. However, much of that is directed at a weaker thesis, which I  call the \emph{weak nomological interpretation}. On that view, the wave function $\Psi$ does not need to be like the classical Hamiltonian to fit into the law system. In particular, $\Psi$ does not need to be  time-independent or simple. This interpretation recommends a weaker criterion for being nomological. The idea is most plausible in some extended Humean framework. In the original Humean framework, laws of nature are the axioms of the best system that summarizes the mosaic. In \cite{loewer2001determinism}, the Humean framework has been extended to allow for deterministic ``chances.'' In \cite{hall2015humean}, it has been further extended to allow intrinsic properties such as mass and charge to be non-fundamental and to be merely part of the best system. 

According to the weak nomological interpretation (Humean version), what is fundamental is just matter (particles or fields) in the four-dimensional spacetime, and the wave function is just a dynamical variable that assists in a simple and informative summary of the mosaic. See \cite{miller2014quantum, esfeld2014quantum, bhogal2015humean, callender2015one}, and \cite{esfeld2017minimalist}.  Although the wave function is part of the best system, it does not need to be time-independent. Moreover, it does not need to be simple \emph{simpliciter}. It just needs to be the simplest one among all competitors. Even though the exact specification of the wave function is complicated, the best system involving the wave function might still be the simplest overall. Albert (p.c.), Maudlin (p.c.), and  \cite{DewarHS} have raised the worry that the complete specification of particle trajectories, which will form another system, seems to postulate much less information than the wave function. This is because the particle trajectories form a single curve in  configuration space, while the wave function assigns values to every point in  configuration space. Moreover, they have raised the worry that the wave function does not supervene on the particle trajectories, since \emph{prima facie} the particle trajectories do not determine the exact values of the wave function. However, it is true that physicists who have access only to facts about the locations of pointers and experimental instruments nonetheless postulate wave functions to make accurate explanations and predictions, and they often agree on the exact wave function of the system. So the best system comparisons and the issue about supervenience may be more complicated than  the debate has assumed. 

At any rate, the weak nomological interpretation  demands less of the wave function of the universe.  It does not need to be a time-independent function,   a simple function, or a function determined in a simple way. It can be time-dependent and highly complex, as long as it is the simplest among all the choices. One could argue that the weak nomological interpretation is less realist than the previous approaches,\footnote{One argument is that the weak nomological interpretation is ``too cheap,'' i.e. too easy to satisfy. We  can presumably extend it to any piece of problematic  ontology and put it in the law system. There is no difficulty, for example, of extending the interpretation to classical physics and move the classical electromagnetic fields from fundamental ontology to the law system. So the fundamental ontology will only consist in particles, not particles plus fields. That would be an unwelcome result, as we can test whether an interpretive framework is realist enough by testing whether  it affirms, in classical field theory, the ontological reality of  fields. } but it could still be realist if the extended Humean framework can be understood as a realist view about laws and properties. So far the weak nomological interpretation has only been defended in the extended Humean framework. It is an open question whether it can be made plausible in some non-Humean framework. 

\section{The \emph{Sui Generis} Interpretation}

It is possible to  be not persuaded by any of the above strategies. The high-dimensional field interpretation and  Hilbert space interpretation require sophisticated stories about the emergence of the apparent three-dimensional objects and ordinary space-time. The low-dimensional multi-field interpretation and the subsystem property interpretation may seem to be trying too hard to squeeze  the wave function into  familiar ontological categories. 

Perhaps the lesson of quantum mechanics is that the wave function does not fit into any familiar categories of things; it is a new kind of entity. Perhaps it is neither ontological nor nomological. In that case, the wave function has its own category of existence that is distinct from anything we have considered. In other words, the wave function is ontologically \emph{sui generis}. \cite{maudlin2013nature} suggests that we should be open to that possibility.

\begin{table}
    \begin{tabular}{ | l | c | c | c | c | c | c | c | c | }
    \hline
    Interpretation & BM & GRW0 & GRWm & GRWf & S0 & Sm & HD & LD  \\ \hline
      High-Dimensional Field & $\cmark$ & $\cmark$ & $\xmark$ & $\xmark$ & $\cmark$ & $\xmark$ & $\cmark$ & $\xmark$  \\ \hline
      Low-Dimensional Multi-field & $\cmark$ & $\cmark$ & $\cmark$ & $\cmark$ & $\cmark$ & $\cmark$ & $\xmark$ & $\cmark$  \\ \hline
      Properties of Systems & $\cmark$ & $\cmark$ & $\cmark$ & $\cmark$ & $\cmark$ & $\cmark$ & $\xmark$ & $\cmark$  \\ \hline
      Vector in Hilbert Space & ? & $\cmark$ & ?  & ?  & $\cmark$ & ? & $\cmark$ & $\xmark$  \\ \hline
      Strongly Nomological  & $\cmark$ & ? & ? & ? & ?  & $\cmark$ & $\cmark$ & $\cmark$  \\ \hline
      Weakly Nomological  & $\cmark$ & ?  & $\cmark$ & $\cmark$ & ?  & $\cmark$ & $\cmark$ & $\cmark$  \\ \hline
       \emph{Sui Generis}  & $\cmark$ & $\cmark$ & $\cmark$ & $\cmark$ & $\cmark$ & $\cmark$ & $\cmark$ & $\cmark$  \\ \hline
    \end{tabular}
\caption{\footnotesize{The first column lists all the realist interpretations  reviewed in this article. In the first row, we have the main solutions to the quantum measurement problem: BM (Bohmian mechanics with a particle ontology), GRW0 (GRW theory without additional ontologies), GRWm (GRW theory with a mass-density ontology), GRWf (GRW theory with a flash ontology), S0 (Everettian theory without additional ontologies), and Sm (Everettian theory with a mass-density ontology).  ``HD'' stands for the view that the fundamental physical space is high-dimensional ($10^{80}$ dimensions in configuration space fundamentalism or possibly infinity in Hilbert space fundamentalism), and ``LD'' stands for the view that the fundamental space is low-dimensional ($3$ dimensions of ordinary physical space). We mark their compatibility with a check  (compatible), a cross  (incompatible), or a question mark (unknown compatibility).}}
\end{table}

\section{Conclusion}
In this article, we have surveyed three kinds of realist interpretations of the wave function:  ontological interpretations,  nomological interpretations, and the \emph{sui generis} interpretation. (See Table 1 for a summary.) A century after the discovery of quantum mechanics, although there is no consensus on what it means, we have made significant progress in constructing several realist interpretations.  Almost every interpretation requires further  developments, and it is too early to say which one is the best or the most fruitful. It is also too early to think that those are exhaustive of all the options available to the realist. In all likelihood, there will be other  ways to think about the wave function from the realist perspective that we have never considered.\footnote{For book-length treatments of the subject, see \cite{ney2013wave}, \cite{lewis2016quantum}, \cite{gao2017meaning}, and \cite{NeyMS}. For clear introductions to the foundations of quantum mechanics, see \cite{bell2004speakable, albert1994quantum, ghirardi2005sneaking, bricmont2016making},  \cite{norsen2017foundations}, and \cite{maudlin2019philosophy}.
For an alternative perspective on realism about the wave function, see  \cite{halvorson2018realist}.}


\section*{Acknowledgement}

I am grateful for useful comments from the editors and two anonymous referees of \emph{Philosophy Compass} and  helpful discussions with David Albert, Valia Allori, Sean Carroll, Christina Conroy, Alexander Ehmann, Sheldon Goldstein, Veronica Gomez,  Mario Hubert, Barry Loewer, Ricardo Martinez, Bradley Monton, Bradley Rettler, Thad Roberts, Davide Romano, and Ashmeet Singh.


\bibliography{test}


\end{document}